\documentclass[a4paper,12pt]{article}
\usepackage{amsmath}
\usepackage{amsthm}
\usepackage{amssymb}
\usepackage{mathrsfs}
\usepackage{graphics}
\usepackage{cite}
\usepackage{indentfirst}
\usepackage{braket}
\usepackage{enumerate}
\usepackage[hidelinks]{hyperref}   
\usepackage{cite}

\DeclareMathOperator{\Tr}{Tr}
\DeclareMathOperator{\spec}{spec}

\renewcommand*{\Re}{\mathop{\mathrm{Re}}\nolimits}
\renewcommand*{\Im}{\mathop{\mathrm{Im}}\nolimits}

\begin{document}

\title {\bf\large Higher-order corrections to the Redfield equation with respect to the system-bath coupling based on the hierarchical equations of motion}

\author{{\normalsize Anton Trushechkin}\footnote{e-mail: trushechkin@mi-ras.ru}}
\date{\normalsize\em Steklov Mathematical Institute of Russian Academy of Sciences, Moscow, Russia\\
National Research Nuclear University MEPhI, Moscow, Russia\\ 
National University of Science and Technology MISIS, Moscow, Russia}
\maketitle
%\firstcollaboration{Submitted by}

%\received{}

\begin{abstract}
The Redfield equation describes the dynamics of a quantum system weakly coupled to one or more reservoirs and is widely used in theory of open quantum system. However, the assumption of weak system-reservoir coupling is often not fully adequate and higher-order corrections to the Redfield equation with respect to the system-bath coupling is required. Here we propose a general method of derivation of higher-order corrections to the Redfield quantum master equation  based on the hierarchical equations of motion (HEOM). Also we derive conditions of validity of the Redfield equation as well as the additional  secular approximation for it.
\end{abstract}

\section{Introduction} 

Since the experimental observations of quantum coherent effects in excitation energy transfer (EET) in photosynthetic light-harvesting complexes \cite{Engel,Scholes}, theoretical description of quantum EET in biological systems attracts much attention (see, e.g., \cite{QEffBio}). This field of research offers also certain challenges for mathematical research. 

One of the mathematical challenges is derivation of quantum master equation governing the dynamics of EET. An EET process in molecular systems can be described from the viewpoint of theory of open quantum systems \cite{BP,Huelga,MayKuhn}. In this case,
 the electronic degrees of freedom of molecules constitute ``a system'', which is coupled to ``a bath'' (or ``a reservoir'') consisting of the vibrational degrees of freedom and the environment of the molecules. Quantum master equations for the  reduced density matrix of the system (i.e., of the electronic degrees of freedom) is a widely used tool of theory of open quantum system. There are well-known rigorous derivations of Markovian quantum master equations for the cases of weak system-bath coupling \cite{Davies,AccLuVol,AccKozLec} and low density of the particles in the bath \cite{Dumke,APV1,APV2} as well as a general form of a generator of a quantum dynamical semigroup -- the so called Gorini--Kossakowski--Sudarshan--Lindblad (GKSL) form \cite{GKS,Lindblad}. Another well-known limiting case is the weak dipole coupling between electronic excitations of molecules. This leads to F\"orster theory of EET \cite{Forster1,Forster2}. The corresponding EET dynamics is also Markovian. Note that, in the physical literature, weak system-bath coupling theory is also referred to as Redfield theory \cite{Redfield}. 
 
Unfortunately, the mentioned limiting regimes leading to Markovian dynamics are not satisfied in light-harvesting complexes. Hence, non-Markovian quantum master equations are required. A possible approach is the derivation of higher-order (non-Markovian) corrections to the known Markovian master equations. For example, the Redfield quantum master equation corresponds to the second-order perturbation theory with respect to the system-bath coupling. General methods using the projection operator technique or cumulant expansion method allow to derive the corrections of an arbitrary order within a given perturbation theory \cite{BP}. In \cite{4thorder}, explicit expressions of the fourth-order corrections to the Redfield equation were obtained.

In this message, we propose another way of derivation of explicit expressions for higher-order corrections to the Redfield equation. This approach is based on the hierarchical equations of motion (HEOM) \cite{KuboTan}. the HEOM is a widely used approach to the non-Markovian dynamics of open quantum systems. The application of this method to the description of EET in molecular systems was presented in \cite{IFl,IFl-FMO}. This method is numerically exact, but computationally expensive, hence, approximate equations based on the HEOM are desired. Note also that the HEOM is suitable only for special types of the spectral density (a function which specifies the system-bath interaction), hence, the same is true for approximate schemes derived from the HEOM.

Another important issue we address in this message is the range of validity of the Redfield equation and the additional secular approximation for it in terms of physical parameters. The mathematical derivations of master equations use formal limits, which are not suitable for physics since they do not give an answer  whether the corresponding limiting case can be applied to a particular physical system or not. The conditions involving the physical parameters are required. In the physical literature, some heuristic conditions of applicability of the Redfield equation in terms of physical parameters are commonly used. An example is the condition that the dipole couplings between electronic degrees of freedom must be much larger than the bath relaxation rates \cite{MayKuhn}. However, as can be seen from numerical experiments. the Redfield equation works fine even when the dipole couplings are small provided that the system-bath couplings are also small with respect to the bath relaxation rates. Moreover, in this case, the Redfield equation can be simplified using the so called local approach \cite{TrushVol,LocGlob1,LocGlob2}.

There are debates concerning the secular approximation for the Redfield equation \cite{Huelga,Ishizaki,Plenio-sec}. The secular approximation involves discarding of  the highly-oscillating terms  from the Redfield equation. This allows to express the equation in the GKSL form, which guarantees the preservation of positivity. But, in many cases, this approximation is inadequate and looses important effects. Note that, recently, other approximations to the Redfield equation leading to equations of the GKSL form were proposed \cite{NonSecEsposito,NonSecGiovan}. In some cases, the aforementioned  local approach also leading to a GKSL equation can be used. 

In this message we propose a general method of derivation of higher-order corrections to the Redfield quantum master equation based on the HEOM. In Sec.~\ref{SecHEOM}, we give the Frenkel exciton Hamiltonian and the HEOM. In Sec.~\ref{SecHigh}, we derive the general formula of higher-order corrections to the Redfield equation. Finally, in Sec.~\ref{SecVal}, we obtain conditions of validity of the Redfield equation and the secular approximation for it using the derived corrections.

\section{Theoretical background}\label{SecHEOM}

The Hamiltonian describing EET processes in molecular aggregates is as follows (the so called Frenkel exciton Hamiltonian) \cite{QEffBio,IFl,IFl-FMO}:

\begin{subequations}
\begin{eqnarray}
&&H=H^{\rm el}+H^{\rm ph}+H^{\text{el-ph}},\\
&&H^{\rm el}=\sum_{j=1}^N\ket j\varepsilon_j\bra j+\sum_{j=1}^N\sum_{k>j}^N(J_{jk}\ket j\bra k+\text{h.c.}),\\
&&H^{\rm ph}=\sum_{j=1}^NH^{\rm ph}_j,\quad
H^{\rm ph}_j=\sum_i
\left(
\frac{p_{ji}^2}{2M_{ji}}+
\frac12M_{ji}\omega_{ji}^2q_{ji}^2
\right),\\
&&H^{\text{el-ph}}=\sum_{j=1}^N V_j\otimes u_j,
\quad V_j=\ket j\bra j,\:
u_j=\sum_iM_{ji}\omega_{ji}^2d_{ji}q_{ji},\label{Equn}
\end{eqnarray}
\end{subequations}
where ``h.c.'' stands for Hermitian conjugate. Also we assume $\hbar=1$. Here  $H^{\rm el}$ is the electronic (system) Hamiltonian: $N$ is the number of monomers (e.g., individual molecules) in the aggregate, $\ket j$ represents the excited electronic state of the $j$th site (molecule) with all other sites being in the ground state,  $\varepsilon_j$ is the electronic excitation energy of the $j$th site, $J_{jk}$ is the dipole Coulombic coupling constant between the $j$th and $k$th sites. These coupling constants are responsible for EET between the sites. Each site $j$ is coupled to its own phononic bath consisting of harmonic oscillators, with $q_{ji}$ and $p_{ji}$ being the position and momentum operators of the $i$th phonon mode of the corresponding bath. The parameters $M_{ji}$ and $\omega_{ji}$ are the mass and frequency of the corresponding mode, and $d_{ji}$ is the displacement of the equilibrium configuration of the mode between the ground and excited electronic states of the site. These displacements $d_{ji}$ play the role of coupling constants between the system (electronic degrees of freedom) and the bath (phononic degrees of freedom). The system-bath interaction manifests itself in the interaction  Hamiltonian $H^{\text{el-ph}}$. 

We consider the initial state (density operator) of both electronic and phononic degrees of freedom of  the form $\rho^{\rm tot}(0)=\rho(0)\otimes\rho^{\rm ph}$, where $\rho(0)$ is the initial electronic density matrix and $\rho^{\rm ph}=e^{-\beta H^{\rm ph}}/\Tr e^{-\beta H^{\rm ph}}$ is the equilibrium phononic state. The  density operator in the interaction representation $\rho^{\rm tot}(t)=U_I(t)\rho^{\rm tot}(0)U_I(t)^\dag$, $U_I(t)=e^{iH_0t}e^{iHt}$, $H_0=H^{\rm el}+H^{\rm ph}$, satisfies the von Neumann equation
\begin{equation}%\label{EqNeumann}
\dot\rho^{\rm tot}(t)=-i[H^{\text{el-ph}}(t),\rho^{\rm tot}(t)],
\end{equation}
where $H^{\text{el-ph}}(t)=e^{iH_0t}H^{\text{el-ph}}e^{-iH_0t}$, $[A,B]=AB-BA$, and (in the following) $\{A,B\}=AB+BA$.

The system-bath interaction is often specified in terms of the spectral density. The spectral density function for the bath coupled to the $j$th site defined as 
\begin{equation}\label{EqJ}
\mathcal J_j(\omega)=\frac\pi2\sum_iM_{ji}\omega_{ji}d_{ji}^2
[\delta(\omega-\omega_{ji})-\delta(\omega+\omega_{ji})].
\end{equation}
Then, the correlation function of the $j$th bath
\begin{equation}%\label{EqCorrdef}
C_j(t)=\Tr\{u_j(t)u_j\rho_j^{\rm ph}\}
\end{equation}
where
$u_j(t)=e^{iH^{\rm ph}_jt}u_je^{-iH^{\rm ph}_jt}$, can be expressed as
\begin{equation}\label{EqCorr}
\begin{split}
C_j(t)&=\int_{-\infty}^{+\infty}
\omega^2\mathcal J_j(\omega)[n_{\rm BE}(\omega)+1]e^{-i\omega t}\,d\omega\\
&=\int_{0}^{\infty}
\omega^2\mathcal J_j(\omega)\coth\frac{\beta\omega}2\cos\omega t\,d\omega
-i\int_{0}^{\infty}
\omega^2\mathcal J_j(\omega)\sin\omega t\,d\omega
\end{split}
\end{equation}
Here $n_{\rm BE}(\omega)=(e^{\beta\omega}-1)^{-1}$ is the Bose--Einstein distribution, $\beta$ is the inverse temperature of the baths.

Consider the Drude--Lorentz spectral density:
\begin{equation}\label{EqDrude}
\omega^2\mathcal J_j(\omega)=
2\lambda_j\frac{\omega\gamma_j}{\omega^2+\gamma_j^2},
\end{equation}
where $\lambda_j$ and $\gamma_j$ are the reorganization energy and Debye (cutoff) frequency of the corresponding bath. The reorganization energy characterizes the strength of the coupling of electronic and phononic degrees of freedom, and the Debye frequency characterizes a time scale of fluctuation of the electronic energy and dissipation of the phonon reorganization energy. 

Following \cite{IFl,IFl-FMO}, we impose a high-temperature condition $\beta\gamma_j\ll1$ for all $j$. Then we can approximate $\coth(\beta\omega/2)$ by $2/(\beta\omega)$ in Eq.~(\ref{EqCorr}) and obtain the expression for the correlation function:
\begin{equation}\label{Eqg}
C_j(t)=\lambda_j\gamma_j
\left(\frac2{\beta\gamma_j}-i\right)
e^{-\gamma_j t}
\end{equation}
for $t\geq0$ and $C_j(-t)=C^*_j(t)$.

Then, in \cite{IFl-FMO}, it is shown that the dynamics of the reduced density operator of the system $\rho(t)=\Tr_R\rho^{\rm tot}(t)$ satisfies the following hierarchical equations of motion (HEOM):
\begin{subequations}\label{EqHEOM0}
\begin{align}
&\dot\rho(t)=
\sum_{j=1}^N\Phi_j(t)\sigma_{\mathbf{e}_j}(t),\label{Eqrho}\\
&\dot\sigma_{\mathbf{n}}(t)=
-\sum_{j=1}^Nn_j\gamma_j
\sigma_{\mathbf{n}}(t)
+
\sum_{j=1}^N
\left[
\Phi_j(t)\sigma_{\mathbf{n}+\mathbf{e}_j}(t)+
\lambda_jn_j\Theta_j(t) \sigma_{\mathbf{n}-\mathbf{e}_j}(t)
\right].\label{EqADO}
\end{align} 
\end{subequations}
Here $\mathbf n=(n_1,\ldots,n_N)\in\mathbb Z_+^N$ (i.e., all $n_j$ are non-negative integers), $\mathbf e_j=(0,\ldots,0,1,0,\ldots,0)$, where 1 on the $j$th position, $\Phi_j(t)$ and $\Theta_j(t)$ are the superoperators defined as $\Phi_j(t)=i[V_j(t),\,\cdot\,]$ and
$$
\Theta_j(t)=i\gamma_j\left(\frac 2{\beta\gamma_j}[V_j(t),\,\cdot\,]-i\{V_j(t),\,\cdot\,\}\right).
$$
The operators $\sigma_{\mathbf n}$ act in the same Hilbert space as $\rho$ (namely, in $\mathbb C^N$) and are referred to as auxiliary density operators (ADOs). Note that Eq.~(\ref{Eqrho}) can be expressed as a particular case of Eq.~(\ref{EqADO}) if we put $\rho(t)=\sigma_{\mathbf 0}(t)$, where $\mathbf 0=(0,\ldots,0)$. Also we put by definition $\sigma_{\mathbf n}(t)\equiv0$ if $\mathbf n$ has negative elements. The initial conditions for the ADOs are $\sigma_{\mathbf n}(0)=0$ for all $\mathbf n$ with positive elements.

Suppose that $\rho(t)$ is of order one, then, from Eq.~(\ref{EqADO}), the magnitude of $\sigma_{\mathbf n}$ is proportional to $\prod_{j=1}^N c_j^{n_j}$, where $$c_j=\lambda_j\gamma_j\sqrt{\left(\frac2{\beta\gamma_j}\right)^2+1}.$$
Hence, the magnitude of $\sigma_{\mathbf n}$ indefinitely increases for large as $|\mathbf n|\equiv n_1+\ldots n_N$ if $c_j>1$ for all $j$, and the last condition cannot be excluded. In \cite{Shi}, a rescaling of ADOs was proposed. For our purposes, it will be convenient to introduce a slightly different rescaling, namely:
\begin{equation}
\tilde\sigma_{\mathbf n}(t)=\left(\prod_{j=1}^N \lambda_j^{n_j}n_j!\right)^{-1}\sigma_{\mathbf n}(t).
\end{equation}
Then the magnitudes of $\tilde\sigma_{\mathbf n}(t)$ are bounded and tend to zero as $|\mathbf n|\to\infty$. Also, since we will study the limiting case of small system-bath couplings, we substitute all $d_{ji}$ in Eq.~(\ref{Equn}) by $\nu d_{ji}$, where $\nu$ is a formal small dimensionless parameter. This is equivalent to the replacement of all $\lambda_j$ by $\nu^2\lambda_j$, see Eqs.~(\ref{EqJ}) and~(\ref{EqDrude}). Then the hierarchy (\ref{EqHEOM0}) is rewritten as
\begin{subequations}\label{EqHEOM1}
\begin{align}
&\dot\rho(t)=
\nu^2\sum_{j=1}^N\lambda_j\Phi_j(t)\sigma_{\mathbf{e}_j}(t),\label{Eqrho1}\\
&\dot\sigma_{\mathbf{n}}(t)=
-\sum_{j=1}^Nn_j\gamma_j
\sigma_{\mathbf{n}}(t)
+
\sum_{j=1}^N
\left[
\nu^2
(n_j+1)\lambda_j\Phi_j(t)\sigma_{\mathbf{n}+\mathbf{e}_j}(t)+
\Theta_j(t) \sigma_{\mathbf{n}-\mathbf{e}_j}(t)
\right],\label{EqADO1}
\end{align} 
\end{subequations}
where we have removed the tildes from the ADOs since, in the following, we will consider only rescaled ADOs.

There are several ways of truncation of the infinite hierarchy of equations (\ref{EqHEOM1}). One commonly accepted way is setting $\sigma_{\mathbf n}(t)\equiv0$ for $|\mathbf{n}|$ larger some threshold $\mathcal N$. According to \cite{IFl,IFl-FMO}, this threshold value should satisfy
\begin{equation*}
\mathcal N\gg\frac{\omega_{\max}}{\min(\gamma_1,\ldots,\gamma_N)},
\end{equation*}
where $\omega_{\max}$ is the largest  difference between the eigenvalues of $H^{\rm el}$. Then, for $|\mathbf{n}|=\mathcal N$, we have
\begin{equation}\label{EqTruncTNL}
\dot\sigma_{\mathbf{n}}(t)=
-\sum_{j=1}^Nn_j\gamma_j
\sigma_{\mathbf{n}}(t)
+
\sum_{j=1}^N
\Theta_j(t) \sigma_{\mathbf{n}-\mathbf{e}_j}(t),
\end{equation}
or,
\begin{equation*}
\sigma_{\mathbf{n}}(t)=\sum_{j=1}^N\int_0^tds\,
e^{-\left(\sum_{k=1}^N\gamma_k n_k\right)s}\,
\Theta_j(t-s) \sigma_{\mathbf{n}-\mathbf{e}_j}(t-s).
\end{equation*}
If we assume that $\sigma_{\mathbf{n}-\mathbf{e}_j}(t-s)$ evolves much slower than $e^{-\left(\sum_{k=1}^N\gamma_k n_k\right)s}\,
\Theta_j(t-s)$ decays (with the increase of $s$), then we can perform the  following approximation:
\begin{equation*}
\sigma_{\mathbf{n}}(t)\cong\sum_{j=1}^N
\left\lbrace
\int_0^tds\,
e^{-\left(\sum_{k=1}^N\gamma_k n_k\right)s}\,
\Theta_j(t-s) 
\right\rbrace
\sigma_{\mathbf{n}-\mathbf{e}_j}(t),
\end{equation*}
and, hence, for $|\mathbf{n}|=\mathcal N-1$,
\begin{equation}\label{EqTruncTL}
\begin{split}
\dot\sigma_{\mathbf{n}}(t)&\cong
-\sum_{j=1}^Nn_j\gamma_j
\sigma_{\mathbf{n}}(t)
+
\sum_{j=1}^N
\Theta_j(t) \sigma_{\mathbf{n}-\mathbf{e}_j}(t)
\\
&+
\sum_{j,k=1}^N
\nu^2
(n_j+1)\lambda_j\Phi_j(t)
\left\lbrace
\int_0^tds\,
e^{-\left(\sum_{l=1}^N\gamma_l n_l+\gamma_j\right)s}\,
\Theta_k(t-s) 
\right\rbrace
\sigma_{\mathbf{n}+\mathbf{e}_j-\mathbf{e}_k}(t).
\end{split}
\end{equation}
So, $\sigma_{\mathbf n}(t)$ for $|\mathbf{n}|=\mathcal N$ do not have to be stored in the computer memory and $|\mathbf{n}|=\mathcal N-1$ is actually the last level of the hierarchy. Equations (\ref{EqTruncTNL}) and (\ref{EqTruncTL}) for the last level of the hierarchy represent two truncation schemes. The first scheme was considered, for example, in \cite{IFl,IFl-FMO}, the second one was considered in \cite{IshiTani,Xu,Shi,Kleine}.

In particular, Eq.~(\ref{EqTruncTL}) for $|\mathbf{n}|=\mathbf{0}$, i.e., for $\sigma_{\mathbf 0}(t)\equiv\rho(t)$, is the Redfield equation (this is a well-known fact and the details will be given in Sec.~\ref{SecVal}):
\begin{equation}\label{EqRedf}
\dot\rho(t)\cong
\sum_{j=1}^N
\nu^2
\lambda_j
\int_0^tds\,
\Phi_j(t)
\Theta_j(t-s) e^{-\gamma_js}
\rho(t).
\end{equation}
This corresponds to the second-order perturbation with respect to $\nu$. Extending the upper limit of integration in Eq.~(\ref{EqRedf}) to infinity corresponds to the Markovian approximation. Correspondingly, the hierarchies (\ref{EqHEOM1}) with the truncation (\ref{EqTruncTL}) for $|\mathbf{n}|\leq1$ provides higher-order corrections to the Redfield equation with respect to the small parameter $\nu$. In the next section, we propose higher-order corrections to the Redfield equation not involving ADOs, i.e. in the form of closed single equations for $\rho(t)$ (like the Redfield equation itself and the fourth-order correction in \cite{4thorder}).

\section{Higher-order  quantum master equations}\label{SecHigh}

Expand the ADOs $\sigma_{\mathbf n}(t)$, $\mathbf{n}\neq\mathbf{0}$, in the formal series with respect to the small parameter $\nu$:
\begin{equation*}%\label{EqExpand}
\sigma_{\mathbf n}(t)=\sum_{m=0}^\infty\nu^{2m}
\sigma^{(m)}_{\mathbf n}(t).
\end{equation*}
The substitution of this expansion into the hierarchy (\ref{EqHEOM1}) and equating the expressions with equal orders of $\nu$ in both sides of Eq.~(\ref{EqADO1}) to each other, gives
\begin{subequations}\label{EqHEOMexpand}
\begin{align}
&\dot\rho(t)=
\sum_{m=1}^\infty\nu^{2m}
\sum_{j=1}^N\lambda_j\Phi_j(t)\sigma^{(m-1)}_{\mathbf{e}_j}(t),\label{Eqrhoexpand}\\
&\dot\sigma^{(m)}_{\mathbf{n}}(t)=
-\sum_{j=1}^Nn_j\gamma_j
\sigma^{(m)}_{\mathbf{n}}(t)
+
\sum_{j=1}^N
\left[
(n_j+1)\lambda_j\Phi_j(t)\sigma^{(m-1)}_{\mathbf{n}+\mathbf{e}_j}(t)+
\Theta_j(t) \sigma^{(m)}_{\mathbf{n}-\mathbf{e}_j}(t)
\right],\label{EqADOexpand}
\end{align} 
\end{subequations}
$m=0,1,2,\ldots,$ with the agreement $\sigma^{(m)}_\mathbf{n}(t)\equiv0$ for $m=-1$. The $2M$th-order quantum master equation corresponds to the neglection of the terms for $m>M$ in Eq.~(\ref{Eqrhoexpand}) and substitution of the functions $\sigma^{(m)}_{\mathbf{e}_j}(t)$  for $0\leq l\leq M-1$ and $j=1,\ldots,N$ by their explicit expressions via $\rho(t)$.

Rewrite Eq.~(\ref{EqADOexpand}) in the form
\begin{equation*}
\sigma^{(m)}_{\mathbf{n}}(t)=\sum_{j=1}^N
\int_0^tds\left[
(n_j+1)\lambda_j\Phi_j(s)\sigma^{(m-1)}_{\mathbf{n}+\mathbf{e}_j}(s)+
\Theta_j(s) \sigma^{(m)}_{\mathbf{n}-\mathbf{e}_j}(s)
\right]e^{-\left(\sum_{k=1}^N\gamma_k n_k\right)(t-s)}.
\end{equation*}
Iterating this expression, we obtain
\begin{multline}\label{EqSigma}
\sigma^{(m)}_{\mathbf{n}}(t)=
\sum_{
\begin{smallmatrix}
{\rm paths}\\
(\mathbf{n};m)\to(\mathbf{0};0)
\end{smallmatrix}
}
\int_0^tds_1\int_0^{s_1}ds_2\ldots\int_0^{s_{2m+|\mathbf{n}|-1}}
ds_{2m+|\mathbf{n}|}
\\
\left[\prod_{l=1}^{2m+|\mathbf{n}|}
{\rm Op}_l(s_l)\right]
e^{-\left(\sum_{k=1}^N\gamma_k n_k\right)t}
\rho(s_{2m+|\mathbf{n}|}),
\end{multline}
where summation is over admissible paths on the lattice
$$\mathbb Z_+^{N+1}=
\{(\tilde n_1,\ldots,\tilde n_N;\tilde m)|\,\tilde n_i\geq0 \text{ for all }i,\,\tilde m\geq0\}$$
starting from $(\mathbf{n};m)$ and ending with $(\mathbf{0};0)$. A path is a sequence of points $(\mathbf{n}^{(l)};m^{(l)})$, $l=0,\ldots,L$, in $\mathbb Z_+^{N+1}$.  Only the following steps from $(\mathbf{n}^{(l-1)};m^{(l-1)})$ to  $(\mathbf{n}^{(l)};m^{(l)})$ are admissible:
\begin{enumerate}[(i)]
\item $\tilde{\mathbf{n}}^{(l)}=\tilde{\mathbf{n}}^{(l-1)}+\mathbf{e}_j$ for some $j$ and $\tilde m^{(l)}=\tilde m^{(l-1)}-1$. Then, in (\ref{EqSigma}), 
$${\rm Op}_l(s)=n^{(l)}_j\lambda_j\Phi_j(s)e^{-\gamma_js}\equiv
n^{(l)}_j\lambda_j\tilde\Phi_j(s);$$
\item $\tilde{\mathbf{n}}^{(l)}=\tilde{\mathbf{n}}^{(l-1)}-\mathbf{e}_j$ for some $j$ and $\tilde m^{(l)}=\tilde m^{(l-1)}$. Then ${\rm Op}_l(s)=\Theta_j(s)e^{\gamma_js}\equiv\tilde\Theta_j(s)$.
\end{enumerate}
Also transitions to the points $(\mathbf{0};\tilde m)$ for $\tilde m>0$ are forbidden.

The end point of the path is $(\mathbf{0};0)$. To complete the path starting from the point $(\mathbf{n};m)$, we have to do overall $m$ steps of type~(i), which reduces the last coordinate by one. But the ``cost'' of each such step is the increase of $|\tilde{\mathbf{n}}|$ by one. Hence, we also have to do overall $m+|\mathbf{n}|$ steps of type~(ii) in order to reduce $|\tilde{\mathbf{n}}|$ to zero. Thus, the length of the path from $(\mathbf{n};m)$ to $(\mathbf{0};0)$ is $L=2m+|\mathbf{n}|$.

The order of integration in Eq.~(\ref{EqSigma}) can be changed to
\begin{equation}\label{EqOrder}
\int_0^t
ds_{2m+|\mathbf{n}|}
\int_{s_{2m+|\mathbf{n}|}}^t
ds_1\int_{s_{2m+|\mathbf{n}|}}^{s_1}ds_2\ldots
\int_{s_{2m+|\mathbf{n}|}}^{s_{2m+|\mathbf{n}|-2}}
ds_{2m+|\mathbf{n}|-1}
\end{equation}

The substitution of Eq.~(\ref{EqSigma}) with order of integration (\ref{EqOrder}) to (\ref{Eqrhoexpand}) gives the formal series
\begin{equation}\label{EqTNL}
\dot\rho(t)=\sum_{m=1}^\infty\nu^{2m}\int_0^tds\,\mathcal K^{(2m)}(t,s)\rho(s),
\end{equation}
where
\begin{subequations}\label{EqTNLker}
\begin{align}
&\mathcal K^{(2)}(t,s)=\sum_{j=1}^N\lambda_j
\tilde\Phi_j(t)\tilde\Theta_j(s),\\
&\mathcal K^{(2m)}(t,s)=
\sum_{
\begin{smallmatrix}
{\rm paths}\\
(\mathbf{0};m)\to(\mathbf{0};0)
\end{smallmatrix}
}
\int_s^tds_1\int_s^{s_2}ds_3\ldots
\int_s^{s_{2m-3}}ds_{2m-2}\,
\prod_{l=0}^{2m-1}
{\rm Op}_l(s_l),\label{EqTNLkergen}
\end{align}
\end{subequations}
$m\geq2$, with the agreement $s_0=t$ and $s_{2m-1}=s$. The paths in (\ref{EqTNLkergen}) start from the point $(\mathbf{0};m)$. The admissible steps are given above. Transitions to the points $(\mathbf{0};\tilde m)$ for $0<\tilde m<m$ are still forbidden.

As we said before, the master equation of order $2M$ corresponds to keeping the first $M$ terms in (\ref{EqTNL}) and neglecting all the remaining terms. 

Equation (\ref{EqTNL}) contains convolution with respect to time and, thus, is time-nonlocal. Let us derive a time-local (convolutionless) equation, see \cite{BP,Huelga,MayKuhn} for general theory. For this aim, we can express
\begin{equation*}
\rho(s)=\rho(t)-[\rho(t)-\rho(s)]=
\rho(t)-\int_s^t dt'\dot\rho(t')=
\rho(t)
-\sum_{m=1}^\infty
\nu^{2m}
\int_s^t dt'\int_0^{t'}ds'\,
\mathcal K^{(2m)}(t',s')\rho(s').
\end{equation*}
Iteration of this equation and the substitution of the result to Eq.~(\ref{EqTNL}) gives
\begin{equation}\label{EqTL}
\dot\rho(t)=\sum_{m=1}^\infty\nu^{2m}\mathcal R^{(2m)}(t)\rho(t),
\end{equation}
where
\begin{subequations}\label{EqTLker}
\begin{multline}
\mathcal R^{(2m)}(t)=
\int_0^tds\,\mathcal K^{(2m)}(t,s)\\-
\sum_{K=1}^{m-1}
\sum_{
\begin{smallmatrix}
m_0,\ldots,m_K\geq1
\\
m_0+\ldots+m_K=m
\end{smallmatrix}
}
\int_0^tds
\int_s^tdt_1
\int_0^{t_1} ds_1
\ldots
\int_{s_{K-1}}^tdt_{K}
\int_0^{t_{K}} ds_{K}
\,
(-1)^{K}\prod_{l=0}^K
\mathcal K^{(2m_l)}(t_l,s_l)
\end{multline}
for $m\geq2$
with the agreement $t_0=t$ and $s_0=s$ and
\begin{equation}
\mathcal R^{(2)}(t)=\int_0^tds\,\mathcal K^{(2)}(t,s).
\end{equation}
\end{subequations}
Again, the time-local master equation of order $2M$ corresponds to keeping the first $M$ terms in (\ref{EqTNL}) and neglecting all the remaining terms. 

Formulas (\ref{EqTNL})--(\ref{EqTNLker}) for the time-nonlocal corrections to the Redfield equation and (\ref{EqTL})--(\ref{EqTLker}) for the time-local (convolutionless) ones are the main result of this section.

Consider the examples. For $m=1$, there are exactly $N$ admissible paths from $(\mathbf{0};1)$ to $(\mathbf{0};0)$:
\begin{equation*}
(\mathbf{0};1)\xrightarrow{\lambda_j\tilde\Phi_j}
(\mathbf{e}_j;0)\xrightarrow{\tilde\Theta_j}
(\mathbf{0};0),\qquad j=1,\ldots,N,
\end{equation*}
where the operators ${\rm Op}$ corresponding to the steps (see Eq.~(\ref{EqSigma})) are indicated above the arrows. Hence, for $M=2$, the Redfield equation (\ref{EqRedf}) is restored.

Consider the fourth-order corrections: $m=2$. The admissible paths from $(\mathbf{0};2)$ to $(\mathbf{0};0)$ are:
\begin{equation*}
\begin{split}
&(\mathbf{0};2)\xrightarrow{\lambda_j\tilde\Phi_j}
(\mathbf{e}_j;1)\xrightarrow{2\lambda_j\tilde\Phi_j}
(2\mathbf{e}_j;0)\xrightarrow{\tilde\Theta_j}
(\mathbf{e}_j;0)\xrightarrow{\tilde\Theta_j}(\mathbf{0};0),\\
&(\mathbf{0};2)\xrightarrow{\lambda_j\tilde\Phi_j}(\mathbf{e}_j;1)\xrightarrow{\lambda_k\tilde\Phi_k}(\mathbf{e}_j+\mathbf{e}_k;0)\xrightarrow{\tilde\Theta_k}
(\mathbf{e}_j;0)\xrightarrow{\tilde\Theta_j}
(\mathbf{0};0),\quad k\neq j,\\
&(\mathbf{0};2)\xrightarrow{\lambda_j\tilde\Phi_j}(\mathbf{e}_j;1)\xrightarrow{\lambda_k\tilde\Phi_k}(\mathbf{e}_j+\mathbf{e}_k;0)\xrightarrow{\tilde\Theta_j}
(\mathbf{e}_k;0)\xrightarrow{\tilde\Theta_k}
(\mathbf{0};0),\quad k\neq j.
\end{split}
\end{equation*}
Hence, 
\begin{equation}\label{EqK4}
\mathcal K^{(4)}(t,s)=
\sum_{j,k=1}^N\lambda_j\lambda_k
\int_s^{t}ds_1\int_s^{s_1}ds_2\,
\tilde\Phi_j(t)\tilde\Phi_k(s_1)
\big[\tilde\Theta_j(s_2)\tilde\Theta_k(s)
+\tilde\Theta_k(s_2)\tilde\Theta_j(s)\big].
\end{equation}
For $N=1$ (single bath), formula (\ref{EqK4}) is reduced to a particular case (for the Drude--Lorentz spectral density (\ref{EqDrude}) and high-temperature approximation) of the fourth-order kernel derived in \cite{4thorder}. Formula (\ref{EqK4}) for $N>1$ 
provides a generalization of such kernel for multibath case. Note that, in contrast to $\mathcal K^{(2)}$, the fourth-order kernel $\mathcal K^{(4)}$ is not reduced to a sum of independent contributions from each bath: the terms with $j\neq k$ describe the interaction between the baths via the system.

Also
\begin{equation}\label{EqR4}
\mathcal R^{(4)}(t)=\int_0^tds\,\mathcal K^{(4)}(t,s)-
\int_0^tds\int_s^tdt'\int_0^{t'}ds'\,\mathcal K^{(2)}(t,s)\mathcal K^{(2)}(t',s').
\end{equation}

Analogously, the sixth-order kernel can be derived:
\begin{equation}%\label{EqK6}
\begin{split}
\mathcal K^{(6)}(t,s)=
\sum_{j,k,l=1}^N\lambda_j\lambda_k\lambda_l
\int_s^{t}ds_1&\int_s^{s_1}ds_2\int_s^{s_2}ds_3
\int_s^{s_3}ds_4\\
\Big\{\tilde\Phi_j(t)\tilde\Phi_k(s_1)\tilde\Phi_l(s_2)
\big[
&\tilde\Theta_j(s_3)\tilde\Theta_k(s_4)\tilde\Theta_l(s)
+\tilde\Theta_j(s_3)\tilde\Theta_l(s_4)\tilde\Theta_k(s)\\
+&\,\tilde\Theta_k(s_3)\tilde\Theta_j(s_4)\tilde\Theta_l(s)
+\tilde\Theta_k(s_3)\tilde\Theta_l(s_4)\tilde\Theta_j(s)\\
+&\,\tilde\Theta_l(s_3)\tilde\Theta_j(s_4)\tilde\Theta_k(s)
+\tilde\Theta_l(s_3)\tilde\Theta_k(s_4)\tilde\Theta_j(s)
\big]\\
+\,\tilde\Phi_j(t)\tilde\Phi_k(s_1)\tilde\Theta_j(s_2)&\tilde\Phi_l(s_3)
\big[
\tilde\Theta_k(s_4)\tilde\Theta_l(s)
+\tilde\Theta_l(s_4)\tilde\Theta_k(s)
\big]\\
+\,\tilde\Phi_j(t)\tilde\Phi_k(s_1)\tilde\Theta_k(s_2)&\tilde\Phi_l(s_3)
\big[
\tilde\Theta_j(s_4)\tilde\Theta_l(s)
+\tilde\Theta_l(s_4)\tilde\Theta_j(s)
\big]\Big\},
\end{split}
\end{equation}
\begin{equation}
\begin{split}
\mathcal R^{(6)}(t)&=
\int_0^tds\,\mathcal K^{(6)}(t,s)-
\int_0^tds\int_s^tdt'\int_0^{t'}ds'\,
[\mathcal K^{(4)}(t,s)\mathcal K^{(2)}(t',s')+
\mathcal K^{(2)}(t,s)\mathcal K^{(4)}(t',s')]
\\&+
\int_0^tds\int_s^tdt_1\int_0^{t_1}ds_1
\int_{s_1}^tdt_2\int_0^{t_2}ds_2
\,\mathcal K^{(2)}(t,s)\mathcal K^{(2)}(t_1,s_1)
\mathcal K^{(2)}(t_2,s_2).
\end{split}
\end{equation}

\section{On the range of validity of the Redfield equation}\label{SecVal}

In this section we discuss the range of validity of the Redfield equation (\ref{EqRedf}) as well as the secular approximation for it. At first, let us express it in a more convenient way. Denote $\spec H^{\rm el}$ the spectrum of $H^{\rm el}$ and
\begin{equation*}
\Omega=\{\varepsilon'-\varepsilon\,\|\,
\varepsilon,\varepsilon'\in\spec H^{\rm el}\}
\end{equation*}
the set of Bohr frequencies, i.e., differences between the eigenvalues of $H^{\rm el}$. Note that if $\omega\in\Omega$, then also $-\omega\in\Omega$. For $\varepsilon\in\spec H^{\rm el}$, denote $P_\varepsilon$ the projector onto the eigensubspace corresponding to $\varepsilon$. Also, for $\omega\in\Omega$, denote 
\begin{equation*}
V_j^{(\omega)}=\sum_{
\varepsilon,\varepsilon':\:
\varepsilon'-\varepsilon=\omega
}
P_\varepsilon V P_{\varepsilon'}, \qquad
\Phi^{(\omega)}_j=i[V_j^{(\omega)},\,\cdot\,],\qquad
\Theta_j^{(\omega)}=i\gamma_j\left(\frac 2{\beta\gamma_j}[V_j^{(\omega)},\,\cdot\,]-i\{V_j^{(\omega)},\,\cdot\,\}\right).
\end{equation*}
Then 
\begin{equation*}
\Phi_j=\sum_{\omega\in\Omega}\Phi_j^{(\omega)},\qquad
\Phi_j(t)=\sum_{\omega\in\Omega}
\Phi_j^{(\omega)}e^{-i\omega t}=
\sum_{\omega\in\Omega}
\Phi_j^{(-\omega)}e^{i\omega t},
\end{equation*}
the same equalities are satisfied for $\Theta_j$ as well.

If we express the terms in (\ref{EqRedf}) as
\begin{equation}\label{EqPhiThOm}
\Phi_j(t)=\sum_{\omega'\in\Omega}\Phi_j^{(-\omega')}e^{i\omega' t},
\qquad
\Theta_j(t-s)=\sum_{\omega\in\Omega}\Theta_j^{(\omega)}
e^{-i\omega(t-s)},
\end{equation}
then, using
$V_j^{(-\omega)}=V_j^{(\omega)\dag}$, where $\dag$ denotes the Hermitian conjugation, the Redfield equation can be rewritten as
\begin{equation}\label{EqNonMark}
\dot\rho(t)=\nu^2
\sum_{j=1}^N
\sum_{\omega,\omega'\in\Omega}
e^{i(\omega'-\omega)t}\,
\Gamma_j(\omega,t)
\left(
V_j^{(\omega)}\rho(t) V_j^{(\omega')\dag}
-
V_j^{(\omega')\dag}V_j^{(\omega)}\rho(t)
\right)
+\text{h.c.},
\end{equation}
where h.c. stands for Hermitian conjugate terms, and
\begin{equation}\label{EqGamma}
\Gamma_j(\omega,t)=\int_0^t C_j(s)e^{i\omega s}\,ds=
\lambda_j\gamma_j
\left(\frac2{\beta\gamma_j}-i\right)
\frac{1-e^{-\gamma_jt+i\omega t}}{\gamma_j-i\omega}.
\end{equation}

For $t\gg\gamma_j^{-1}$, we can take the limit $t\to\infty$ in Eq.~(\ref{EqGamma}) and equation (\ref{EqNonMark}) is reduced to an autonomous matrix differential equation:
\begin{equation}\label{EqMark}
\dot\rho(t)=\nu^2
\sum_{j=1}^N
\sum_{\omega,\omega'\in\Omega}
e^{i(\omega'-\omega)t}\,
\Gamma_j(\omega)
\left(
V_j^{(\omega)}\rho(t) V_j^{(\omega')\dag}
-
V_j^{(\omega')\dag}V_j^{(\omega)}\rho(t)
\right)
+\text{h.c.},
\end{equation}
where
\begin{equation}\label{EqGammaInf}
\Gamma_j(\omega)=\int_0^\infty C_j(s)e^{i\omega s}\,ds
=
\frac{\lambda_j\gamma_j}{\gamma_j-i\omega}
\left(\frac2{\beta\gamma_j}-i\right).
\end{equation}
If the time scale of evolution of $\rho(t)$ is much larger than $\max_j\gamma_j^{-1}$, then we can replace $\Gamma_j(\omega,t)$ by $\Gamma(\omega)$ in (\ref{EqNonMark}) for all times: on small times, where the finiteness of the limit of integration in Eq.~(\ref{EqGamma}) is essential, $\rho(t)$ does not succeed to  evolve significantly and the introduced error is small. In other words, equation (\ref{EqMark}) can be used for the description of evolution of $\rho(t)$ for all $t\geq0$.

The solution of Eq.~(\ref{EqMark}) can be regarded as an action of a semigroup $\{\Lambda_t\}_{t\geq0}$: $\rho(t)=\Lambda_t\rho(0)$. The semigroup property $\Lambda_s\Lambda_t=\Lambda_{s+t}$ represents the Markovian property of the dynamics. Equation (\ref{EqMark}) is referred to as the Markovian Redfield equation. Accordingly, equation (\ref{EqNonMark}) is often referred to as the non-Markovian Redfield equation, but it differs from Eq.~(\ref{EqMark}) only on small times (of order $\max_j\gamma_j^{-1}$).

Unfortunately, the semigroup generated by Eq.~(\ref{EqMark}), generally speaking, does not preserve the positivity: the condition of positive semidefiniteness of $\rho(t)$ may be not satisfied even if this condition is satisfied for $\rho(0)$. To fix this drawback, a further approximation is commonly made. Equation (\ref{EqMark}) contains oscillating terms proportional to $e^{i(\omega'-\omega)t}$ for $\omega'\neq\omega$. If the time scale of evolution of $\rho(t)$ is much larger than $\max|\omega'-\omega|^{-1}$, where the maximum is taken over $\omega,\omega'\in\Omega$, $\omega\neq\omega'$, then the terms with $\omega\neq\omega'$ can be neglected as rapidly oscillating. This approximation is called the secular approximation. Then Eq.~(\ref{EqMark}) is reduced to
\begin{equation}\label{EqSec}
\dot\rho(t)=-i[H_{\rm LS},\rho(t)]+
\nu^2
\sum_{j=1}^N
\sum_{\omega\in\Omega}
2\Re[\Gamma_j(\omega)]
\left(
V_j^{(\omega)}\rho(t) V_j^{(\omega)\dag}
-
\frac12\{V_j^{(\omega)\dag}V_j^{(\omega)},\rho(t)\}
\right),
\end{equation}
where
$$
H_{\rm LS}=
\nu^2
\sum_{j=1}^N
\sum_{\omega\in\Omega}
\Im[\Gamma_j(\omega)]
V_j^{(\omega)\dag}V_j^{(\omega)}
$$
is the Lamb shift Hamiltonian. Equation (\ref{EqSec}) has the Gorini--Kossakowski--Sudarshan--Lindblad (GKSL) form and, hence, preserves the positivity (\cite{GKS,Lindblad}, see also \cite{BP,Huelga}). Equation (\ref{EqSec}) is referred to as the secular Redfield equation. 

Now we discuss the range of validity of the Redfield equation (\ref{EqRedf}) (or (\ref{EqNonMark})) as well as its Markovian and secular versions (\ref{EqMark}) and (\ref{EqSec}) in terms of physical parameters rather than formal dimensionless parameter $\nu$. So, we put $\nu=1$ since this is a formal parameter and  all information about the system-bath interaction is given in the physical parameters like $\lambda_j$, $\gamma_j$, etc.

The time scale of evolution of $\rho(t)$ in Eqs.~(\ref{EqNonMark})--(\ref{EqSec}) is, roughly, $\min_{j,\omega}|\Gamma_j(\omega)|^{-1}$, where
\begin{equation}\label{EqGammabs}
|\Gamma_j(\omega)|=
\frac{\lambda_j\gamma_j}{\sqrt{\gamma_j^2+\omega^2}}
\sqrt{\left(\frac{2}{\beta\gamma_j}\right)^2+1}.
\end{equation}
Note that we work in the high-temperature approximation $\beta\gamma_j<1$ and, thus, expression (\ref{EqGammabs}) can be simplified to $|\Gamma_j(\omega)|=2\lambda_j\gamma_j/\beta\sqrt{\gamma_j^2+\omega^2}$. However, we keep the more general expression (\ref{EqGammabs}) trying to extrapolate the results to the low-temperature case as well. According to the previous discussion, the range of validity of the Markovian approximation is 
$\max_{j,\omega}|\Gamma_j(\omega)|\ll\min_j\gamma_j$, or
\begin{equation}\label{EqCondMark}
\max_{
\begin{smallmatrix}
j\in\{1,\ldots,N\}
\\
\omega\in\Omega
\end{smallmatrix}
}\frac{\lambda_j\gamma_j}{\sqrt{\gamma_j^2+\omega^2}}
\sqrt{\left(\frac{2}{\beta\gamma_j}\right)^2+1}
\ll
\min_{j\in\{1,\ldots,N\}}\gamma_j.
\end{equation}
Also the range of validity of the secular approximation is
\begin{equation}\label{EqCondSec}
\max_{
\begin{smallmatrix}
j\in\{1,\ldots,N\}
\\
\omega\in\Omega
\end{smallmatrix}
}\frac{\lambda_j\gamma_j}{\sqrt{\gamma_j^2+\omega^2}}
\sqrt{\left(\frac{2}{\beta\gamma_j}\right)^2+1}
\ll
\min_{
\begin{smallmatrix}
\omega,\omega'\in\Omega
\\
\omega\neq\omega'
\end{smallmatrix}
}
|\omega-\omega'|.
\end{equation}

It remains to establish the range of validity of the second-order approximation
\begin{equation*}
\dot\rho(t)=\mathcal R^{(2)}(t)\rho(t)=
\int_0^tds\,
\mathcal K^{(2)}(t,s)\rho(t),
\end{equation*}
 which was the starting point for all equations of this section.
To estimate its range of validity, let us analyze the fourth-order corrections:
\begin{equation}\label{EqFourth}
\begin{split}
\dot\rho(t)&=\mathcal R^{(2)}(t)\rho(t)
+\mathcal R^{(4)}(t)\rho(t)
\\&=
\int_0^tds\,
\mathcal K^{(2)}(t,s)
\Big[
\rho(t)-\int_s^tdt'\int_0^{t'}ds'\,
\mathcal K^{(2)}(t',s')\rho(t)
\Big]+\int_0^tds\,
\mathcal K^{(4)}(t,s)\rho(t)
\end{split}
\end{equation}
If the fourth-order corrections are small, we can conclude that the second-order approximation is adequate. Of course, this is not a rigorous proof, but rough estimates. The second term in the square brackets (the double integral  with $\mathcal K^{(2)}$) represents the evolution of $\rho$ from $s$ to $t$ in the second-order approximation. The influence of the second term in the square brackets is negligible if the evolution of $\rho$  in the second-order approximation is much slower than the decay of $\mathcal K^{(2)}(t,s)$. Since the $\mathcal K^{(2)}(t,s)$ decays as $\exp[-\min_j\gamma_j(t-s)]$, and the rate of evolution of $\rho$ is given by $\max_{j,\omega}|\Gamma(\omega)|$, we again arrive at the condition (\ref{EqCondMark}).

Consider not the last term of (\ref{EqFourth}):

\begin{multline*}
\int_0^tds\,
\mathcal K^{(4)}(t,s)\rho(t)\\=
\sum_{j,k=1}^N\lambda_j\lambda_k
\int_0^tds_1\int_0^{s_1}ds_2\int_0^{s_2}ds_3
\,
\tilde\Phi_j(t)\tilde\Phi_k(s_1)
\big[\tilde\Theta_j(s_2)\tilde\Theta_k(s_3)
+\tilde\Theta_k(s_2)\tilde\Theta_j(s_3)\big]\rho(t).
\end{multline*}
For rough estimation of this term, consider the case of large $t$, namely, $t\gg\max_j\gamma_j^{-1}$. Then, using the representation (\ref{EqPhiThOm}), we obtain
\begin{multline*}
\int_0^\infty ds\,
\mathcal K^{(4)}(t,s)\rho(t)=
\sum_{j,k=1}^N
\sum_{\omega_1,\omega_2,\omega_3,\omega_4\in\Omega}
\left[
\frac{
\Phi_j^{(-\omega_1)}
\Phi_k^{(-\omega_2)}
\Theta_j^{(\omega_3)}
\Theta_k^{(\omega_4)}
}
{
\gamma_j-i\omega_4
}
+
\frac{
\Phi_j^{(-\omega_1)}
\Phi_k^{(-\omega_2)}
\Theta_k^{(\omega_3)}
\Theta_j^{(\omega_4)}
}
{\gamma_k-i\omega_4}
\right]
\\
\times
\frac{
e^{i(\omega_1+\omega_2-\omega_3-\omega_4)}
}
{
[\gamma_j+\gamma_k-i(\omega_4+\omega_3)]
[\gamma_j-i(\omega_4+\omega_3-\omega_2)]
}\rho(t)
\end{multline*}
The magnitude of each term is
\begin{multline}\label{EqMagn4}
\frac{2(\lambda_j\gamma_j)(\lambda_k\gamma_k)}{\sqrt{\gamma_j^2+\omega_4^2}}
\sqrt{
\left\lbrace\left(\frac{2}{\beta\gamma_j}\right)^2+1\right\rbrace
\left\lbrace\left(\frac{2}{\beta\gamma_k}\right)^2+1\right\rbrace
}\\
\times
\left\{\left[(\gamma_j+\gamma_k)^2+(\omega_4+\omega_3)^2\right]
\left[\gamma_j^2+(\omega_4+\omega_3-\omega_2)^2\right]\right\}^{-1/2}
.
\end{multline}
We can estimate the last factor in (\ref{EqMagn4}) from above by putting $\omega_4+\omega_3=0$ and $\omega_4+\omega_3-\omega_2=0$. Then, expression  (\ref{EqMagn4}) is much smaller than the magnitude of the second-order contribution (\ref{EqGammabs}) (for $\omega=\omega_4$) whenever
\begin{equation}\label{EqCondSecond}
\max_{k\in\{1,\ldots,N\}}\lambda_k
\sqrt{\left(\frac{2}{\beta\gamma_k}\right)^2+1}\ll
\min_{j\in\{1,\ldots,N\}}\gamma_j.
\end{equation}
We can see that this is a more stringent version of condition (\ref{EqCondMark}).

Thus, condition (\ref{EqCondSecond}) is a (rough) sufficient condition for the validity of the Redfield equation (both Markovian and non-Markovian versions), and (\ref{EqCondSec}) is a sufficient condition of the secular approximation. This two inequalities are main results of this section.

\section{Conclusions}

We have derived explicit general formulas for the corrections to the Redfield equation of arbitrary order with respect to the system-bath coupling based on the HEOM: formulas  for the time-nonlocal equation (\ref{EqTNL})--(\ref{EqTNLker}) and time-local (convolutionless) one (\ref{EqTL})--(\ref{EqTLker}). For the considered particular case (see below), the fourth-order corrections coincide with the  corrections previously derived in the framework of the projection operator method \cite{4thorder}. For high orders, the explicit expressions of the equations become cumbersome and, hence, not practically useful. However, the existence of a general scheme which allows to obtain closed equations for the density matrix (without auxiliary density operators) of arbitrary order seems to be of some interest.

In this work, we adopted the HEOM for the special case of the Drude--Lorentz spectral density (\ref{EqDrude}) and the high-temperature approximation. Also we considered a particular system-bath Hamiltonian. In this case, the bath correlation function is an exponent (\ref{Eqg}). In general, the HEOM is applicable to the case when the correlation function is a sum of exponents. The proposed method of derivation of the corrections to the Redfield equation works for this case as well. Theoretically, the exponential decomposition for the bath correlation functions can be applied for the general case, but it may be computationally inefficient.

Recently, a HEOM based on an alternative a decomposition scheme for the bath correlation functions was proposed \cite{CHEOM}. Namely, the decomposition scheme in \cite{CHEOM} is based on the Chebyshev polynomials and Bessel functions. The corresponding HEOM is referred to as the C-HEOM (Chebyshev-HEOM). It is suitable for the cases where the exponential decomposition is inefficient, but has its own shortcomings. It would be interesting to derive corrections to the Redfield equations based on the C-HEOM.

Also we have derived conditions of the validity of the Redfield equation (Ineq.~(\ref{EqCondSecond})) as well as the secular approximation (Ineq.~(\ref{EqCondSec})) for it in terms of physical parameters. The derivation of Ineq.~(\ref{EqCondSecond}) is still heuristic, but has more solid basis that the heuristic conditions commonly used in the physical literature. Rigorous derivation of such conditions is still required. Note that, in \cite{Tere}, the validity of the Redfield equation and the secular approximation is studied for the model which is exactly solvable by the pseudomode method \cite{Tere0}. A heuristic derivation of the range of validity of the F\"orster and modified Redfield theories is proposed in \cite{TrushJCP}.

\textbf{Acknowledgments.}  This work was supported by the Russian Science Foundation (project 17-71-20154).

\end{document}